\documentclass[prb,twocolumn,superscriptaddress]{revtex4-2}
\usepackage{amsfonts}
\usepackage{amssymb}
\usepackage{amsmath}
\usepackage{graphicx}
\usepackage{dcolumn}
\usepackage{bm}
\usepackage{units}
\usepackage{soul}
\usepackage{multirow}
\usepackage{CJKutf8}
\usepackage{subfigure}
\usepackage{color}%
\usepackage{url}
\usepackage[colorlinks,linkcolor=blue,anchorcolor=blue,citecolor=blue,urlcolor=blue]{hyperref}
\usepackage{array}
\usepackage{ulem}
\begin{document}
\title{A Unified Theory of Unusual Anisotropic Magnetoresistance and Unidirectional Magnetoresistance in Nanoscale Bilayers}
\author{X. R. Wang}
\email{phxwan@ust.hk}
\affiliation{School of Science and Engineering, Chinese University of Hong Kong (Shenzhen), Shenzhen 51817, China}

\affiliation{Department of Physics, The Hong Kong University of Science and Technology, Clear Water Bay, Kowloon, Hong Kong, China}

\begin{abstract}
Nanoscale bilayers containing at least one magnetic layer exhibit universal unusual anisotropic magnetoresistance (UAMR) 
and unidirectional magnetoresistance (UMR). They are currently understood through various mechanisms related to the 
interconversion of charge current and spin current, giant magnetoresistance, thermal magnonic effects, thermoelectric effects, 
and diverse spin-dependent scattering processes. This raises a fundamental question: do the universal behaviors observed 
in a wide range of systems stem from underlying general principles? We demonstrate here that both UAMR and UMR arise 
from electron transport influenced by the magnetization vector present in the magnetic material and the interfacial potential 
inherent in heterostructures. Specifically, UAMR represents current-independent resistance (resistivity) of bilayers. 
UMR is the resistance proportional to the current although electron transports of the bilayers are the linear response 
to high current densities and their induced thermal gradients. Our theory introduces a novel approach that considers 
the interplay between the magnetization vector, thermal gradients, and the effective internal electric field at the interface. 
This framework provides a unified explanation for both UMR and UAMR, effectively capturing key experimental features 
such as dependence on current direction, magnetization orientation, film thickness, and magnetic field strength. 
Furthermore, it offers a universal perspective that bridges UMR and UAMR effects, enhancing our understanding of 
spin-dependent transport phenomena in bilayers.
\end{abstract}
\maketitle

The study of magnetoresistance in nanoscale multilayers has gained significant attention due to its fundamental interest 
and important applications in spintronic devices. The discovery of the giant magnetoresistance (MR) and the tunneling 
MR \cite{Grunberg,Fert,Miyazaki,Moodera,Yuasa,Parkin}  opened the field of spintronics near the end of the last century. 
Studies of magnetotransport in recent years have given birth to many new concepts and phenomena, such as spin 
Seebeck \cite{Uchida}, spin pumping \cite{Gerrit}, spin Hall MR (SMR) \cite{Gerrit1,Gerrit2}—termed as unusual 
anisotropic MR (UAMR) in this study—and unidirectional MR (UMR) \cite{Gambardella1, Gambardella2,Gambardella3}.
Among these, UAMR and UMR have emerged as key phenomena that reflect the intricate interplay between charge 
transport and magnetism.

UAMR is the resistance of bilayers at low current density and depends on the directional magnetization perpendicular 
to the current. For a current density $\vec{j} \parallel \hat{x}$, flowing in the plane of a ferromagnet-nonmagnet 
(FM/NM) bilayer, the longitudinal and transverse resistivity depends on the magnetization direction $\vec{m}$ as:
$\rho_{xx} = \rho_0 + \rho_1m_x^2 + \rho_2m_z^2 + \rho_3m_z^4 + \rho_4m_x^2m_z^2$ and 
$\rho_{xy} = \rho_5m_z + \rho_6m_z^3 + \rho_1m_xm_y + \rho_4m_xm_ym_z^2$.
This is in contrast to the conventional anisotropic MR (AMR), which is given by $\rho_{xx} = \rho_0 + \rho_1m_x^2$
 and $\rho_{xy}=\rho_5m_z + \rho_1m_xm_y$. Here, the parameters $\rho_i$ (where $i=0,1,\ldots,6$) are 
independent of the magnetization direction.

This newly observed UAMR has stimulated the development of the SMR theory \cite{Gerrit,Gerrit1,Gerrit2}, which 
attributes UAMR to spin/charge current interconversion and spin current transmission/reflection at bilayer interface. 
SMR theory predicts $\rho_{xx} = \rho_0 + \rho_1m_x^2 + \rho_2m_y^2$ which has been widely used to interpret 
UAMR in bilayers \cite{MR1,MR2,MR3,MR4,MR5,MR6,MR7}. The theory is also used to understand UMR 
\cite{Gambardella1,Gambardella2,Gambardella3}, spin-torque ferromagnetic resonance \cite{STFMR1,STFMR2}, 
harmonic Hall voltage \cite{Hhall}, magnetic field sensing \cite{sensor}, and magnetization or Néel-vector switching 
\cite{switching1,switching2,switching3,switching4,switching5}. UAMR is a ubiquitous phenomenon observed in all 
kinds of nanoscale multilayers made of metals, insulators, and semiconductors with strong, weak, or negligible 
spin-orbit interactions, as well as in systems with topological or non-topological states. Clearly, the original SMR 
theory cannot apply to this wide range of multilayers. As a result, alternative spin-current-related MR models have 
been proposed to explain the universal nature of UAMR. These include Rashba-Edelstein MR \cite{REMR1,REMR2}, 
spin-orbit MR \cite{SOMR}, anomalous Hall MR \cite{AHMR}, orbital Hall MR \cite{OMR1}, orbital Rashba-Edelstein 
MR  \cite{OMR2}, Hanle MR  \cite{OMR3}). A natural question arises: Why does the same UAMR emerge from so 
many different microscopic origins? Why do so many different microscopic interactions exhibit the same angular 
dependence? Wouldn't it be more natural to expect a unified origin for UAMR? The question becomes more urgent 
if one notices many inherent issues in MR-like theories \cite{xrw1,xrw2,xrw3} and predicted $\rho_{xx}$ not 
accurately aligning with most experimental observations. Indeed, a recent two-vector theory predicts exactly the 
observed $\rho_{xx}$ and $\rho_{xy}$ of UAMR \cite{xrw1,xrw2,xrw3}.  In this two-vector theory, the key 
components are magnetization of ferromagnetic (FM) layers and the effective internal interfacial electric field 
arising at the interface between ferromagnetic (FM) and non-magnetic (NM) materials, influencing the behavior 
of charges and spins. These two ingredients result in the universal behaviors of UAMR in FM/NM bilayers and FM 
single layers, independent of the details of microscopic interactions among electrons, lattices, spins, and magnons.

UMR is the nonlinear MR of bilayers at high current density [above $10^{10}$A/m$^2$] \cite{Gambardella1,
Gambardella2,Gambardella3,BUMR1,BUMR2,BUMR3,BUMR4,BUMR5,BUMR6,BUMR7,UMR3,UMR4,UMR5,UMR6,UMR7,
UMR8,UMR9}. Different from UAMR whose resistivity does not depend on the current and follows $\rho_{xx}=\rho_0
+\rho_1m_x^2$ when $\vec m$ rotates in the $xy$-plane, $\rho_{xx}$ in UMR is linear in current density and is odd 
in $m_y$. Both $\rho_{xx}$ and $\rho_{xy}$ are sensitive to the magnetic field, film thickness, and temperature. 
Under an alternating current of frequency $\omega$, single harmonic resistivity of $\rho_{xx}^\omega$ and 
$\rho_{xy}^\omega$ follow the behaviour of UAMR while the second harmonic resistivity of $\rho_{xx}^{2\omega}$ 
and $\rho_{xy}^{2\omega}$ reverses sign with the directions of current and transverse field, and has a strong 3-fold 
symmetry at  low magnetic field and only a 1-fold symmetry of $\rho_{xx}^{2\omega}\propto m_y$ and 
$\rho_{xy}^{2\omega}\propto m_x$ at high magnetic field. The same behaviour is observed in a wide range of systems, 
including FM/NM \cite{Gambardella1,Gambardella2,Gambardella3,BUMR1,BUMR2,BUMR3,BUMR4},  FM/semiconductor 
\cite{BUMR5}, FM/topological insulator bilayers \cite{BUMR6,BUMR7}, and 2D materials \cite{UMR9}.  
Its microscopic origin is still under debate. To date, UMR is attributed to multiple origins including giant MR effect, spin 
Hall effect and inverse spin Hall effect, the Rashba-Edelstein effect, magnon effect, spin-flip and spin-dependent 
scatterings, and thermal effects. Considering coexistence of UAMR and UMR in vast different systems and two vector 
theory can account all behaviors of UAMR \cite{xrw3}, it is natural to ask whether two-vector theory can also 
explain the universal features of UMR, and this is the main theme of this study.

In this paper, we propose a unified theoretical framework that accounts for both UAMR and UMR in nanoscale bilayers. 
We show that the underlying physics of universal features of UAMR and UMR are electron transport under the two 
vectors: magnetization and interfacial field.
\begin{figure}[htbp]
\centering
\includegraphics[width=8.5cm]{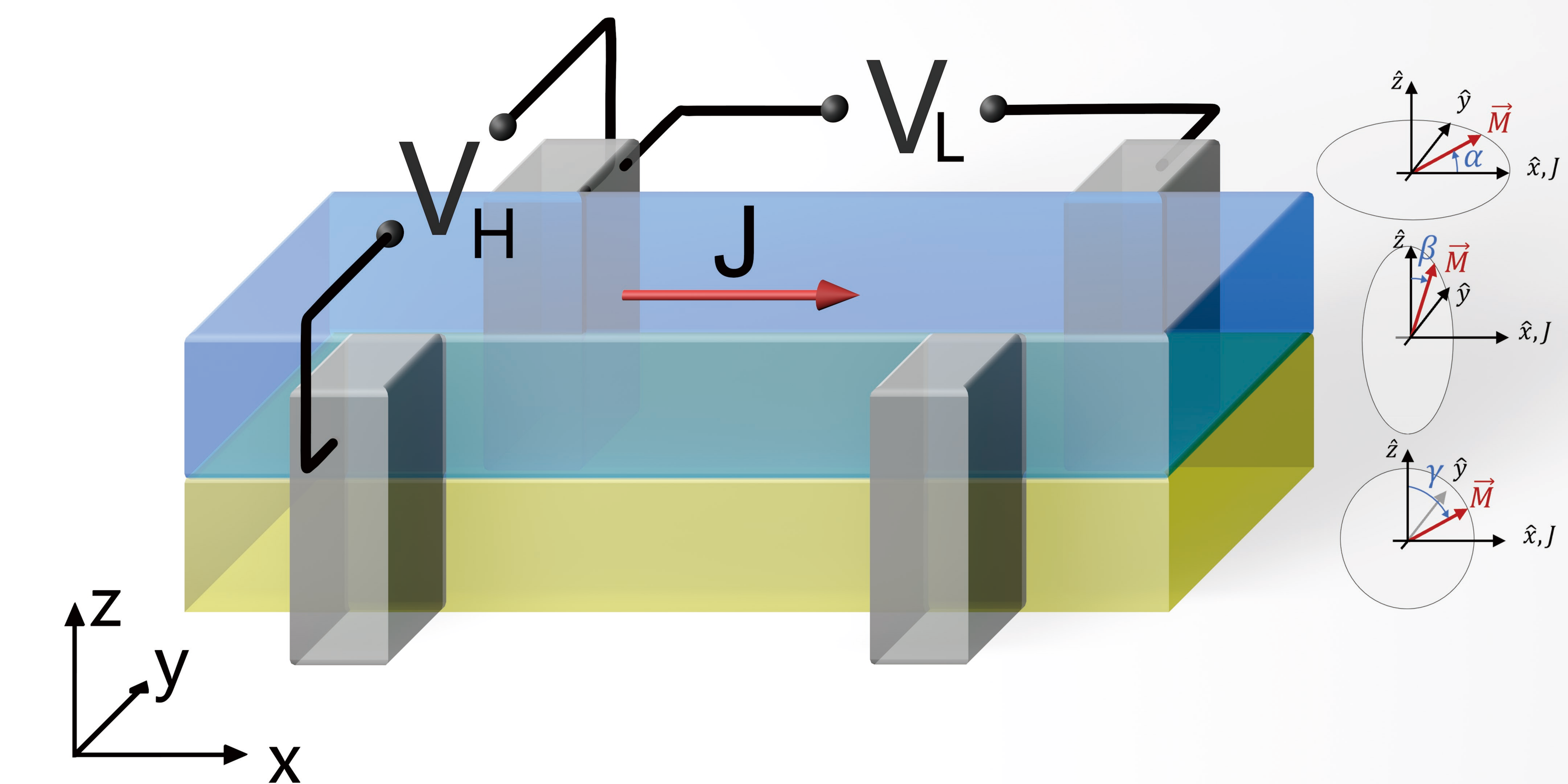}\\
\caption{Illustration of experimental set-up of a bilayer laying in the $xy$-plane. The blue is a FM layer and the yellow 
is a NM layers. Both of them are nanometer thick. They are either metal or non-metal, topological or non-topological, 
and can even be 2D FM and NM materials. The current flows along the $x$-direction. From the measurments of $V_L$ 
and $V_H$ and device geometry, one can obtain $\rho_{xx}$ and $\rho_{xy}$. $\alpha$ is the angle between $\vec M$ 
and the $x$-axis when $\vec M$ rotates in the $xy$-plane while $\beta$ and $\gamma$ are the angles between 
$\vec M$ and the $z$-axis when $\vec M$ rotates in the $yz$- or $zx$-plane indicated in the right coordinates. 
Under a high current density, a thermal gradient will bulid up perpendicular to the bilayer (the $z$-direction).}
	\label{fig1}
\end{figure}

To be  more specific, a bilayer of one FM and a nonmagnetic (NM) metallic layer is consider here although it could be 
various combinations of materials, including ferromagnets (FM), collinear ferrimagnets, collinear antiferromagnets and 
2D magnets. Both layers are nanometer thick, allowing electrons in the metallic layer(s) to experience the magnetization 
($\vec M=M\vec m$) of the FM layer, regardless of whether it is a metal or an insulator, due to quantum tunneling effects. 
The electric field response $\vec E$ to an applied current density $\vec j$ flowing along the $x$-direction of a bilayer in 
the $xy$-plane, as illustrated in Fig. \ref{fig1}, should depend on the magnetization. Additionally, an interfacial field 
($\vec n\parallel \hat z$), which is perpendicular to the interface and arises from chemical potential differences between 
the two layers, will also influence electron motion in the bilayer. This interfacial field should decay exponentially away 
from the interface. Under a high current density $\vec j$ (larger than $10^10A/m^2$) \cite{Gambardella1,Gambardella2,
Gambardella3}, a thermal gradient $\nabla T=a j^2\hat z$ perpendicular to the bilayer is generated due to Joule heating, 
where $a$ is a device parameter. Since the thermal gradient is already a higher-order effect of the applied current, we 
will limit the thermal electric contribution to the electric field in the bilayer to the level of the Nernst effect. 
The most general expression for $\vec E$, which is linear in both $\vec j$ and $\nabla T$, is given by \cite{xrw4,xrw5,xrw6} 
\begin{equation}
\vec E=C_0\nabla T+ C_1\vec m\times \nabla T  +\tensor{\rho}(\vec m, \vec n) \vec j, 
\label{e-field}\end{equation}
where $C_0$ is the Seebeck coefficient and $C_1$ measures the anomalous Nernst effect. 
Thermodynamic quantities must be functions of state variables of which $\vec m$ and $\vec n$ are the only vectors. 
Thus, rank-2 resisitivty tensor $\tensor{\rho}(\vec m, \vec n)$ can only be constructed from $\vec m$ and $\vec n$. 
Generalizing the approach from our earlier publication \cite{xrw1,xrw2} to the current case, and keeping in mind 
that $\nabla T\parallel \vec n\perp \vec j$, the linear response of Eq. \ref{e-field} under a direct current $\vec j$ 
leads to the longitudinal and transverse resistivity: 
\begin{equation}
\begin{aligned}
&\rho_{xx}\equiv \vec E\cdot \hat x/j=\rho +Cjm_y +A_1 m_x^2+A_2m_z \\
&\rho_{xy}\equiv \vec E\cdot\hat y/j= -Cjm_x+A_1 m_xm_y+B_1m_z +B_2, 
\end{aligned}
\label{rho}
\end{equation}
where $C\equiv C_1a$, $\rho$, $A$'s and $B$'s are device parameters and can only depend on $m_z$ \cite{xrw1}. 
Without interfacial field, $A_2=B_2=0$ and all other parameters do not depend on magnetization direction. 
$A_1$ describes usual AMR and planar Hall effect and comes from external electric filed mediated 
electron-magnon scattering, and $B_1$ is the usual anomalous Hall effect. In the presence of an interfacial 
field, extra angular dependence of $\rho_{xx}$ and $\rho_{xy}$ can come from electron-magnon scattering 
mediated by $\vec n$, manifesting through $m_z$ dependences of parameters. $C$, $\rho$, $A_1$, and $B_1$ 
must be even functions of $m_z$ while $A_2$, $A_3$, $B_2$, and $B_3$ must be odd for achiral materials. 
The higher-power terms of these parameters in $m_z$ are negligible because the power in $m_z$ is the 
number of magnons involved in the scatterings. Since thermal electric effect is normally small, we shall keep 
$C$ independent of $m_z$. If one defines $\rho=\rho_0\sum_{n=1}^{n=\infty}d_nm_z^{2n}$; 
$A_1=\rho_1+\rho_4m_z^2+\sum_{n=2}^{n=\infty}a_{1n}m_z^{2n}$; 
$B_1=\sum_{n=0}^{n=\infty}b_{1n}m_z^{2n}$; $A_2=\sum_{n=0}^{n=\infty}a_{2n}m_z^{2n+1}$; 
$B_2=\sum_{n=0}^{n=\infty}b_{2n}m_z^{2n+1}$; $B_3=\sum_{n=0}^{n=\infty}b_{3n}m_z^{2n+1}$; and keep 
the expansions up to ${\vec m}^4$, then the longitudinal and transverse resistivities take following universal forms
\begin{equation}
\begin{aligned}
&\rho_{xx} = \rho_0 + \rho_1m_x^2 + \rho_2m_z^2 + \rho_3m_z^4 + \rho_4m_x^2m_z^2+Cjm_y \\
&\rho_{xy} = \rho_5m_z + \rho_6m_z^3 + \rho_1m_xm_y + \rho_4m_xm_ym_z^2-Cjm_x,
\end{aligned}
\label{UAMR}
\end{equation}
where angular-independent parameters $\rho_2\equiv (d_1+a_{21})$, $\rho_3\equiv (d_2+a_{22})$, 
$\rho_5\equiv (b_{10}+b_{20})$, and $\rho_6\equiv (b_{11}+b_{21})$. $\rho_i  (i=2,3,4,6)$ are from 
$\vec n$-mediated magnon process. Without $C$ terms, this is exactly the behavior of UAMR 
observed in all kinds of multilayers including bilayers and magnetic nanoscale single layers. 
Interestingly, although Eq. \ref{e-field}  is the linear response to external forces, the resistivity appears 
to be nonlinear and propotional to the current density. 
The $C$-term explains well different resistances of Pt/Py bilayer when $m_y$ \cite{BUMR2} or $j$ 
\cite{BUMR4} is reversed at high current density ($>10^{11}\sim10^{12}$A/m$^2$).   

To separate UMR from UAMR, recent experiments \cite{Gambardella1,Gambardella2,Gambardella3,UMR3,UMR4,
UMR5,UMR6,UMR7,UMR8,UMR9} utilize alternating current (AC) $\vec j=j\cos (\omega t)\hat x$ of frequency 
$\omega$ to probe magnetoresistance. By employing either frequency-locking techniques \cite{Gambardella1} 
or the Wheatstone bridge method \cite{UMR5}, the first and second harmonic longitudinal and Hall voltages, 
denoted as $V_L^\omega$,  $V_L^{2\omega}$, $V_H^\omega$,  and $V_H^{2\omega}$ (see Fig. 1) can 
be measured. From them, resistances, $R_{xx}^\omega\equiv V_L^\omega/j$, $R_{xx}^{2\omega}\equiv 
V_L^{2\omega}/j$, $R_{xy}^\omega\equiv V_H^\omega/j$, and $R_{xy}^{2\omega}\equiv V_H^{2\omega}/j$
are obtained. When the magnetization vector $\vec m$ rotates in the $xy$-plane at an angle $\alpha$ with 
respect to $\hat x$, universal behaviors of $R_{xx}^{2\omega}=r_1j(\sin\alpha -\sin 3\alpha)+r_2j\sin\alpha$ 
and $R_{xy}^{2\omega}=r_3j(\cos\alpha -\cos 3\alpha)+r_4j\cos\alpha$ are exactly what were observed  
across all experiments to date, while $R_{xx}^\omega$ and $R_{xy}^\omega$ exhibit UAMR behavior. 
We aim to demonstrate that Eq. \ref{e-field} can not only explain the universal second harmonic resistivity, 
in addition to the UAMR for the first hamonic resistivity, but also predicts field-independent $r_2$ and $r_4$, 
and negligible small $r_1$ and $r_3$ for strong magnetic field and thick bilayers.

At a high alternating current density $\vec j=j\cos (\omega t)\hat x$, an oscillating Oersted field  along 
the $y$-direction $\vec h=dj\cos (\omega t)\hat y$ is generated in addition to a larger static field $H$ 
that is used to control the magnetizationdirection $\vec m$, where $d$ is the thickness of the bilayer. 
Since $\omega$ is typically less than tens of Hz in most experiments, which is significantly lower than 
the GHz frequency of spin response, $\vec m$ should always align with the total effective field and 
oscillates around its averaged direction with frequency $\omega$.  

The polar and azimuthal angles of $\vec m$ are $\theta+ bj\cos (\omega t)\cos\theta\sin\phi$ and 
$\phi+ bj\cos (\omega t)\cos\phi$, where $\theta$ and $\phi$ represent the time-averaged values, 
and $b$ relates to bilayer thickness and the dynamic susceptibility of the FM layer, which is inversely 
proportional to the effective magnetic field \cite{Han,yin}. Thus, we can approximate $m_x(t)$,  $m_y(t)$, 
and $m_z(t)$ as follows: $m_x(t)=\sin[\theta+bj\cos(\omega t)\cos\theta\sin\phi]\cos[\phi+ bj\cos 
(\omega t)\cos\phi]\simeq m_x[1-bj\cos(\omega t)\sin\phi]+bj\cos (\omega t)\cos^2\theta\sin\phi\cos\phi$, 
$m_y(t)=\sin[\theta+bj\cos(\omega t)\cos\theta\sin\phi]\sin[\phi+ bj\cos (\omega t)\cos\phi]\simeq m_y+
bj\cos (\omega t)\sin\theta\cos^2\phi+bj\cos (\omega t)\cos^2\theta\sin^2\phi$, and 
$m_z(t)=\cos[\theta+ bj\cos (\omega t)\cos\theta\sin\phi]\simeq m_z-bj\cos (\omega t)
\sin\theta\cos\theta\sin\phi$. To simplify notation, we denote $m_x$,  $m_y$, and $m_z$ as 
the time-averaged values of $m_x(t)$,  $m_y(t)$, and $m_z(t)$, respectively.
\begin{figure}[htbp]
\centering
\includegraphics[width=8.5cm]{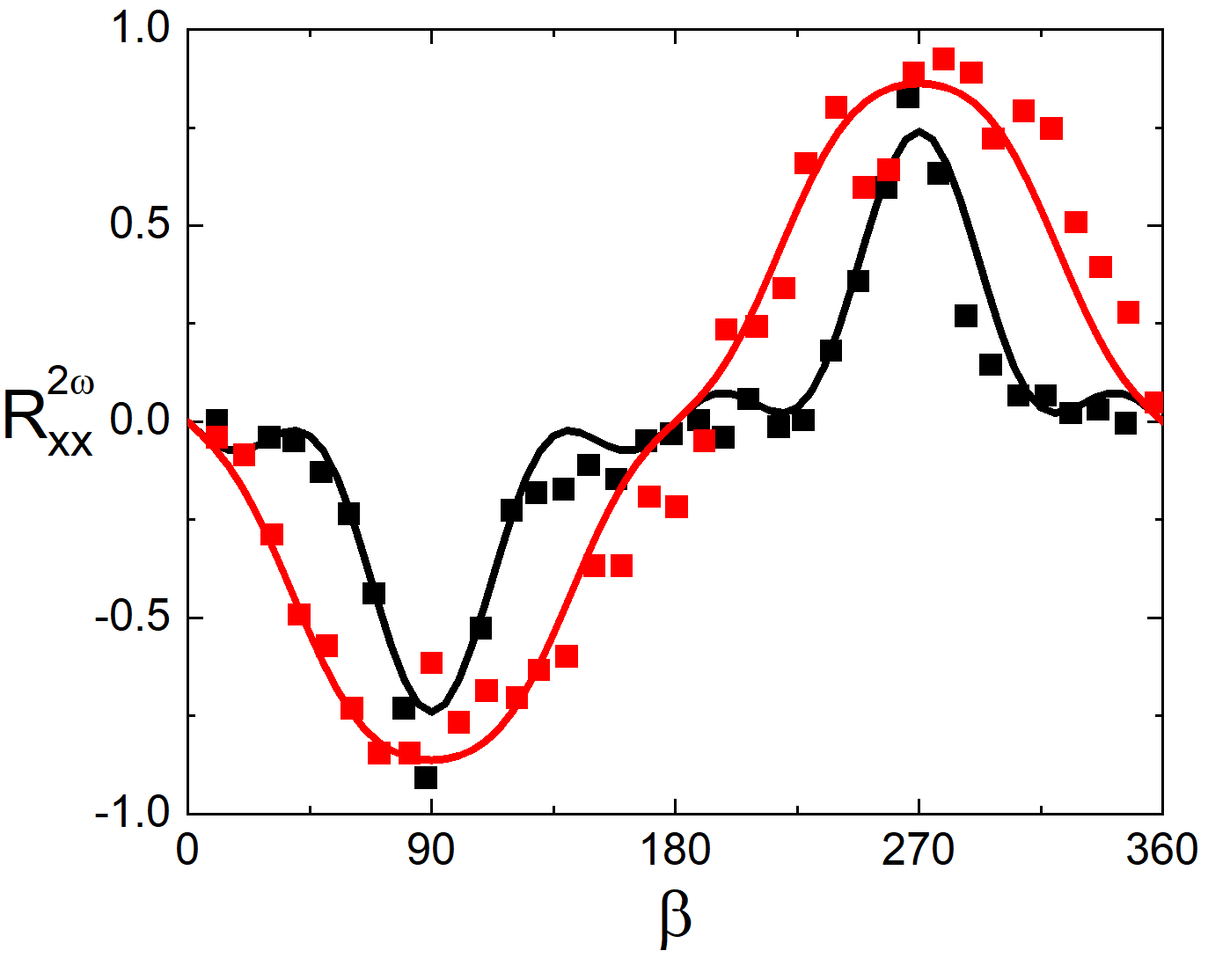}\\
\caption{Experimental data of $R_{xx}^{2\omega}(\beta)$ (symbols) and fits to $R_{xx}^
{2\omega} (\beta)=a_1\sin\beta-a_2(\sin3\beta-\sin\beta)-a_3(\sin5\beta-\sin3\beta)$. 
Experimental data are from Fig. 2b of Ref. \cite{BUMR3} for Pt(4 nm)/Co(2 nm) (the red dots) 
and Pt(4 nm)/Co(1 nm) (the black dots) with $j=10^{11}$A/m$^2$, $B=2$T, and $\omega=801$Hz. 
The fitting parameters are $a_1=-0.75 (-0.31)$m$\Omega$, $a_2j=-0.057 (-0.086)$m$\Omega$, 
$a_3j=-0.041 (0.14)$m$\Omega$ for Pt(4 nm)/Co(2 nm) (Pt(4 nm)/Co(1 nm)).}
	\label{fig2}
\end{figure}

From Eq. \ref{e-field} and $\tensor{\rho}(\vec m, \vec n)$ established earlier \cite{xrw2}, it follows 
straightforwardly that $\rho_{xx}^\omega\equiv E_x^{\omega}/j$ and $\rho_{xy}^\omega\equiv 
E_y^{\omega}/j$ are consistent with Eq. \ref{UAMR}. Below, we examine how $\rho_{xx}^{2\omega}
\equiv E_x^{2\omega}/j$ and $\rho_{xy}^{2\omega}\equiv E_y^{2\omega}/j$ depend on angle 
$\alpha$ when $\vec m$ rotates in the $xy$-plane, as well as on angles $\beta$ and $\gamma$ 
when $\vec m$ rotates in the $yz$- and $zx$-planes, respectively. After some algebra, we obtain:   
\begin{equation}
\begin{aligned}
&\rho_{xx}^{2\omega} (\alpha)=(C+c_1)j\sin\alpha -c_1 j\sin 3\alpha, \\
&\rho_{xy}^{2\omega}(\alpha) =-(C+c_1)j\cos\alpha+c_1j\cos 3\alpha, \\
&\rho_{xx}^{2\omega} (\beta)=(C+c_2)j\sin\beta+(c_3-c_2) j\sin3\beta-c_3j\sin5\beta \\
&\rho_{xy}^{2\omega}(\beta) =c_4j \sin2\beta+c_5j\sin4\beta, \\
&\rho_{xx}^{2\omega} (\gamma)=0, \\
&\rho_{xy}^{2\omega}(\gamma) =c_1+c_6-Cj \sin\gamma-c_1\cos2\gamma-c_6\cos4\gamma, 
\label{UMR1}
\end{aligned}
\end{equation}
where $c_1=\rho_1 b/2$, $c_2=b(\rho_2+\rho_3)/2$,  $c_3=b\rho_3/4$, $c_4=-b(\rho_5/2+3\rho_6/4)$, $c_5=
-3b\rho_6/8$ and $c_6=b\rho_4/8$. $\rho_i$ ($i=2,\ldots 6$) and, in turn, $c_i$  ($i=2,\ldots 6$) come from 
$\vec n$-mediated electron-magnon scattering and should decay exponentially with the layer thickness. 
In other words, the UAMR and UMR are mainly interfacial phenomena, and $\rho_i$'s and $c_i$'s are negligibly 
small for thick enough bilayers. Also $c_i$ ($i=1,\ldots 6$) are proportional to the dynamic suceptibility of FM 
layer which is inversely proportional to the effective magnetic field. 
Thus, all $c_i$-terms are negligibly small in strong magnetic field. However, $C$-terms in Eqs. \ref{UAMR} and 
\ref{UMR1} are due to the Joule heating of high current density that does not depend on the bilayer thickness 
and magnetic field. Although formula for $\rho_{xx}^{2\omega} (\alpha)$ and $\rho_{xy}^{2\omega} (\alpha)$ 
have been obtained from SMR theory and magnonic effects \cite{Gambardella1,Gambardella2,Gambardella3,UMR7}, 
all existing theories have no prediction about $\rho_{xx}^{2\omega}$ and $\rho_{xy}^{2\omega}$ when $\vec m$ 
rotates in the $yz$- and $zx$-plane, and very few measurements have been reported about longitudinal and 
transverse resistance (resistivity) as a function of magnetization direction in the $yz$-  and $zx$-planes. 
To test this theory against all other existing theories, we read $R_{xx}^{2\omega} (\beta)$ [proportional to 
$\rho_{xx}^{2\omega} (\beta)$] from Fig. 1b in Ref. \cite{BUMR3} and fit them by the formula in Eq. \ref{UMR1}. 
The red and black dots in Fig. \ref{fig2} are the experimmental data for Pt(4 nm)/Co(2 nm) (red) and 
Pt(4 nm)/Co(1 nm) (black) with $j=10^{11}$A/m$^2$, $B=2$T, and $\omega=801$Hz. The curves are the fit to 
$R_{xx}^{2\omega} (\beta)=a_1\sin\beta-a_2(\sin3\beta-\sin\beta)-a_3(\sin5\beta-\sin3\beta)$ with fitting 
parameters of $a_1=-0.75 (-0.31)$m$\Omega$, $a_2j=-0.057 (-0.086)$m$\Omega$, $a_3j=-0.041 (0.14)
$m$\Omega$ for Pt(4 nm)/Co(2 nm) (Pt(4 nm)/Co(1 nm)). It is clear that the experimental data supports the 
current theory.

The current theory predicts that the amplitudes of 1-fold and 3-fold symmetries in $\rho_{xx}
^{2\omega} (\alpha)$ are the same as the corresponding amplitudes in $\rho_{xy}^{2\omega} (\alpha)$.
This prediction is consistent with all existing experiments \cite{Gambardella1,Gambardella2,Gambardella3,UMR3,
UMR4,UMR5,UMR6,UMR7,UMR8,UMR9}. The UMR beahves differently when $\vec m$ rotates in the $yz$-plane: 
$\rho_{xx}^{2\omega}(\beta)$ is the linear combination of 1-fold, 3-fold, and 5-fold sinusoidal 
functions while $\rho_{xy}^{2\omega}(\beta)$ is the sum of 2-fold and 4-fold sinusoidal functions. 
When $\vec m$ rotates in the $zx$-plane, $\rho_{xx}^{2\omega} (\gamma)$ is zero and $\rho_{xy}^{2\omega} 
(\gamma)$ has a nonzero mean, 1-fold sinusoidal term, together with a 2-fold and a 4-fold consine terms. 
All these predictions are new, and can be used to test and to disinguish this theory from others.  

It is important to emphasize that the second harmonic signals, which are proportional to the current density,  
result from a linear response  to the current and thermal gradient, nor a nonlinear one. The observed nonlinear 
behavior arises from thermal effects and current-dependent, time-dependent UAMR, as discussed above and in 
Eq. \ref{e-field}. The expressions $\rho_{xx}^{2\omega} (\alpha)$ and $\rho_{xy}^{2\omega} (\alpha)$ are exactly 
what were observed in all UMR experiments \cite{Gambardella1,Gambardella2,Gambardella3,BUMR1,BUMR2, BUMR3,BUMR4,BUMR5,BUMR6,BUMR7,UMR3,UMR4,UMR5,UMR6,UMR7,UMR8,UMR9}. Notably, Eq. \ref{UMR1} 
is the most general form of linear response of electron transport to a magnetization and an internal effective 
field perpendicular to a bilayer. In fact, if one assume magnetization is the only vector in the complete set of 
stater variable as SMR theory assumed, then it is mathematically impossible to admit SMR formula for resisitivity. 
The universal forms of UMR do not depend on microscopic interactions, such as sd-interaction or Rashba and 
Dresselhaus spin-orbit interactions. However, the thickness dependences of $c_i$ ($i=2,\ldots,6$) reveal the 
property of the interfacial field, and the external magnetic-field dependences of $c_i$ tell us the magnetic 
suceptibility of the magnetic layer. It is important to emphaize that expoential decay of UAMR in SMR-like 
theories is attributed to spin diffusion, not the interfacial field-mediated electron-magnon scattering. 
Thus, current theory predicts a stronger UAMR and UMR at a higher temperature, in agreement with 
experiments \cite{UMR8}, while SMR theories would predict a weaker one, opposite to the expoeriments. 
At high fields, $c_i$ ($i=1,\ldots,6$) approach zero, and $\rho_{xx}^{2\omega} $ and $\rho_{xy}^{2\omega}$ 
become $\rho_{xx}^{2\omega} (\alpha)=Cj\sin\alpha$, $\rho_{xy}^{2\omega}(\alpha) =-Cj\cos\alpha$,  
$\rho_{xx}^{2\omega} (\beta)=Cj\sin\beta$, $\rho_{xy}^{2\omega}(\beta) =0$,  $\rho_{xx}^{2\omega}
 (\gamma)=0$, $\rho_{xy}^{2\omega}(\gamma) =-Cj\cos\gamma$, controlled by the same coefficient $C$. 

This work demonstrates that both UAMR and UMR can be understood within a unified theoretical framework. 
The interplay between magnetization, thermal gradients, and interfacial electric fields provides a comprehensive 
explanation for the observed phenomena. This new understanding enables the identification of materials with strong 
UMR effects, particularly in 2D systems and topological materials, facilitating robust UMR signals at room temperature 
for practical device applications. Furthermore, it allows for the integration of UMR with other spintronic effects, 
such as spin-orbit torque, the spin Hall effect, or magnetoelectric effects, to create multifunctional devices.
For the future research, it is important to have a formulation based on microscopic Hamiltonian. Only with such a 
formulation one may know how to use microscopic interactions to manipulate the coefficients in UAMR and UMR.

Our findings suggest that one can select materials with the desired work function difference to control UAMR 
and UMR, and utilize a gate voltage to adjust the interfacial potential, which in turn can tune UAMR and UMR. 
UMR is considered promising for spintronic devices, especially in non-volatile memory (e.g., magnetic 
random-access memory, MRAM), magnetic field sensors, and spin-based logic devices. This work lays the 
foundation for further exploration of spin-dependent transport in multilayers and its implications for future 
spintronic applications. For instance, it should enhance our ability to electrically detect magnetization direction 
and add functionality to spintronic devices.

In conclusion, we have presented a novel theoretical approach to understanding UAMR and UMR. 
According to this theory, UAMR and UMR do not depend on microscopic interactions and do not require 
the interconversion between spin current and charge current. However, if such an interconversion can 
influence electron transport in a system described by only two vectors, then such a microscopic process 
should lead to the predictions presented here. In principle, materials with any spin-electron scattering under 
the influence of an interfacial potential should exhibit the universal UAMR and UMR discussed herein. 
Several predictions are new. Some of them are supported by existing experiments while the others 
are awaited for experimental confirmation.

\begin{acknowledgements}
I thank Dr. Xuchong Hu for drawing the figures. This work is supported by the University Development 
Fund of the Chinese University of Hong Kong, Shenzhen, the National Key Research and Development 
Program of China (No. 2020YFA0309600), the NSFC Grant (No. 12374122),  and Hong Kong RGC Grants 
(No. 16300523, 16300522, and 16302321). 
\end{acknowledgements}

\clearpage
\onecolumngrid

\end{document}